\documentstyle[aasms4,amstex,amsfonts,epsfig,rotating,float]{article}

\lefthead{Fromerth \& Melia}
\righthead{Determination of Central Masses in AGNs}

\begin{document}
\centerline{Submitted to the Astrophysical Journal}
\bigskip
\title{Determination of the Central Mass in AGNs Using Cross-Correlation Lags and Velocity Dispersions}

\author{Michael J. Fromerth$^*$\altaffilmark{1}}
\affil{$^*$Physics Department, The University of Arizona, Tucson, AZ 85721}

\and 

\author{Fulvio Melia$^{\dag}$\altaffilmark{2}}
\affil{$^{\dag}$Physics Department and Steward Observatory, The University of Arizona, Tucson, AZ 85721}

\altaffiltext{1}{NSF Graduate Fellow.}
\altaffiltext{2}{Sir Thomas Lyle Fellow.}

\begin{abstract}
We here estimate the mass $M$ of the central object in five Active Galactic Nuclei (AGNs) using 
the most recent reverberation data obtained by the AGN Watch consortium.
The cross-correlation function (CCF) centroids of the broad Ly${\alpha}$~$\lambda 1216$ 
and \ion{C}{4}~$\lambda 1549$ lines are used to estimate the size of the broad line region 
(BLR) in these sources.  We calculate the velocity dispersions of these lines in the root mean 
square (rms) spectra, and then use our results to estimate $M$.  We argue that our technique 
of calculating the velocity dispersion should work in the general case of an arbitrary line 
profile, unlike methods that depend on the measurement of the full-width at half-maximum 
(FWHM) of the broad line.  We also show that our results agree with the FWHM method in the 
limit of a normal (Gaussian) line profile.  The masses calculated here are considerably
smaller than those calculated with the previous generation of reverberation data.
\end{abstract}


\section{INTRODUCTION}

With our rapidly expanding knowledge of the Galactic Center (see, e.g., Melia 1999), we 
now have a real hope of unraveling how the BLRs in AGNs are produced, and how they evolve 
spatially and temporally.  An understanding of the structure of the BLR in turn is important 
in determining how the gaseous environment behaves near the central engine, presumably 
a supermassive black hole (SMBH), in these sources. Unlike the Galactic Center, however,
where we know the mass very accurately (e.g., \cite{GEOE97}; \cite{Ghez98}), we do not
yet have a reliable estimate of $M$ for these objects.  But since all of the relevant
length scales depend on the central mass, one may reasonably expect that a careful modeling
of these environments with temporally resolved line profiles can provide us with a tightly
constrained range of values for $M$. 

Reverberation methods (e.g., \cite{BM82}; \cite{JC91}) have become popular tools for 
estimating the BLR size in AGNs.  The cross-correlation function (CCF) centroid 
$\tau_{cent}$ for a given line, typically measured in days, represents a delay 
between a change in the continuum flux and a corresponding (presumably causal) change 
in the line flux.  From it, we can infer the responsivity-weighted mean radius, 
$r \sim c \tau_{cent}$, of the line-emitting gas, assuming the delay is due to 
light-travel time effects (\cite{KG91a}).  If we make the very reasonable assumption 
that the central mass dominates the motion of the BLR gas, the velocity dispersion 
$\sigma^2$ of this line can then be used to estimate the central mass, $M \approx 
f c \tau_{cent} \sigma^2 / G$, where $f$ is a corrective factor of order unity 
representing our uncertainty in the geometry and isotropy of the BLR.

An example of how this works in a simplified case, where the velocity distribution of 
the gas is such that the line profile in the root mean square (rms) spectrum assumes 
a roughly Gaussian shape, has recently been studied by Peterson \& Wandel (1999).
Using the UV and optical data for NGC~5548 from the extensive monitoring campaigns 
conducted by the AGN Watch consortium (see, e.g., \cite{KTK95}; \cite{BMP99}), they 
computed the rms spectrum from the set of individual spectra.  The line profiles in 
the rms spectrum represent those portions of the emission lines that are actually varying in response to the continuum fluctuations, and therefore correspond to the CCF measurements.
Using this method, the FWHM of each line is measured directly from the line profile, and 
a velocity dispersion is calculated according to the relation $\sigma^2 = 3 V_{\rm FWHM}^2/ 4$.
Using the reported values of $\tau_{cent}$, they obtained a central mass in this
source of $(6.8 \pm 2.1) \times 10^7 M_{\odot}$.

However, although estimating the velocity dispersion from the FWHM is valid in the case of a Gaussian or other regular (e.g., Laplacian) line profile, this method cannot be used in the general case where the line profile is highly irregular.
In this paper, we introduce a more generally applicable
method to calculate the velocity dispersion directly, and show that our results agree with the 
FWHM procedure in the limit of a Gaussian profile.  We avoid the simplifying assumptions concerning 
the line shape, which allows us to significantly broaden the class of objects for which the
central mass may be estimated using reverberation methods.

\section{METHODOLOGY}
We have assembled and analyzed the UV data obtained with AGN Watch observations of five AGNs:  
3C~390.3 (\cite{PTO98}), Fairall~9 (\cite{PMRP97}), NGC~3783 (\cite{GAR94}), NGC~5548 during 
the $HST$ FOS campaign (\cite{KTK95}), and NGC~7469 (\cite{IW97}).  The first object in this 
list is classified as a broad-line radio galaxy, whereas the others are Seyfert~1 galaxies.
The mean and rms spectra corresponding to each data set have been determined according
to the procedure described in Rodriguez-Pascual, et al. (1997), and are shown for each of the
five sources individually in Figures 1 and 2.  These spectra have already appeared in the
references cited above, but we reproduce them here as a class to demonstrate the range of
line profiles one typically encounters with this process, and also to show the wavelength 
band used for each of the lines in our analysis (see below).  We have selected the 
Ly${\alpha}~\lambda 1216$ and \ion{C}{4}~$\lambda 1549$ lines for our analysis because they 
typically have the strongest signal-to-noise ratios.  In those cases where the (redshifted) 
Ly${\alpha}$ line has been contaminated by geocoronal emission, we have attempted to set the 
lower bound of the wavelength bins beyond the contamination region.  However, the Ly${\alpha}$ 
lines in NGC~5548 and NGC~7469 still show significant contamination (see Fig. \ref{fig:2spectra}).
In all these objects, the Ly${\alpha}$ line is further contaminated by \ion{N}{5}~$\lambda 1240$, 
but no attempt to deblend these two lines has been made here.  The net effect of this is to 
increase our estimate of the Ly${\alpha}$ velocity dispersion.  These contaminants make the 
Ly${\alpha}$ measurements much less reliable than the \ion{C}{4} measurements, which are 
only slightly affected by, e.g., \ion{He}{2}~$\lambda 1640$.

These broad line profiles are understood to be produced by the composite emissivity of a
broad line region (BLR) comprised of distinct clouds, which constitute a distribution 
in velocity around the central mass concentration.  Along any given line-of-sight, say
the $z$-axis, the velocity dispersion of these emitters is defined as
\begin{equation}
\sigma_z^2 \equiv <v_z^2> - <v_z>^2\;,
\label{eq:sigma_z}
\end{equation}
where clearly $v_z$ denotes the $z$-component of the total velocity vector $\vec v$.
Then, assuming that $|\vec v| \ll c$, we can approximate the Doppler-shift formula as 
$(\lambda/\lambda_0) \simeq 1 + v_z / c $, where $\lambda_0$ is the wavelength in
the emitter's frame.  Thus, as long as the velocity field is isotropic (i.e., $\sigma^2 = 3 
\sigma_z^2$), this relation can be used to replace $v_z$ in Equation~(\ref{eq:sigma_z}), 
and thereby to obtain an expression for $\sigma^2$ in terms of observable quantities:
\begin{equation}
\sigma^2 = \frac{3 c^2}{\lambda_0^2}\ (<\lambda^2> - <\lambda>^2)\;.
\label{eq:sigma}
\end{equation}
The mean values of $\lambda^2$ and $\lambda$ appearing on the right hand
side of this expression are often calculated from the data assuming that the
line profiles are Gaussian, for which a determination of the FWHM is then
sufficient.  To keep this procedure as general as possible, we here
instead calculate the expectation (or mean) value $<g>$ of any given function
$g(\lambda)$ of $\lambda$ using a flux-weighted sum: 
\begin{equation}
<g> = \frac{\sum_i g(\lambda_i) F_{\rm line}(\lambda_i)}{\sum_{i} F_{\rm line}(\lambda_i)}\;,
\label{eq:expectation}
\end{equation}
where $F_{\rm line}(\lambda)$ is the (continuum-subtracted) line flux at wavelength 
$\lambda$ and the summation is taken over the appropriate wavelength bins $\lambda_i$.
The uncertainty in the calculated value of $\sigma$ is then estimated by assigning the 
highest and lowest plausible values for the underlying continuum.

In our calculation of $M$, we have used the values of $\tau_{cent}$ published in the references cited
above for the 5 sources considered here.  The uncertainty in the value of $\tau_{cent}$ is estimated 
using the Monte Carlo techniques described in White \& Peterson (1994); however, rather than using
a thin shell model for the BLR, we have instead assumed an extended BLR consistent with the 
observations and the general inferences drawn from modeling (see, e.g., \cite{JB97}).
We chose the line response function to be $\Psi(r, \theta) = (1 + A \cos{\theta}) \sin{\theta}$ 
over the range of radii $R_{in} < r < R_{out}$, where $A$ is the cloud emission anisotropy and 
$R_{in}$ and $R_{out}$ are the inner and outer radial extents, respectively, of the BLR.
For example, a $\Psi(r, \theta)$ that is fairly constant over all BLR radii is consistent with the density 
and velocity fields required to model the \ion{C}{4} $\lambda 1549$ line in Fairall 9 (\cite{FM99}).
The inner extent of the BLR is set to correspond with the \ion{He}{2} $\lambda 1640$ cross-correlation 
lag (typically having the shortest lag of any of the broad emission lines), and the outer extent is 
optimized such that a histogram of modeled line delays peaks at the observed line lag (see \cite{WP94}).
We used the values of $A \sim 1$ for Ly$\alpha$ and $A\sim 0.7$ for \ion{C}{4}~$\lambda 1549$ 
calculated by Ferland, et al. (1992).

Table~\ref{tbl-1} summarizes the results of these calculations.  Our estimates of the errors 
associated with $\tau_{cent}$ are much higher than those obtained using a thin-shell BLR model.
This is intuitively correct, since we would expect an extended BLR to show a much broader, flatter CCF.
In most cases, our extended BLR model also yields larger error estimates for $\tau_{cent}$ than the model-independent method described in Peterson, et al. (1998), which estimates errors based on uncertainties due to the observational sampling and the flux measurements only.
This suggests that the latter method underestimates the actual uncertainty if the BLR is indeed spatially extended.
The sources of error in $\sigma$ include the uncertain fitting of the continuum level, the blending of 
the broad line wings with the continuum, and line contamination (as discussed above).
Since only the former is included in our error estimates, the $\sigma$ errors we quote here
are probably underestimated.  We also restate that the values of $M$ listed in Table~\ref{tbl-1} 
are subject to corrective factors of order unity, stemming from our assumption of an isotropic 
velocity field.

The last column in Table~\ref{tbl-1} lists the masses estimated by Wandel, et al. (1999) for the same objects using H$\beta$ reverberation data.
Their results assume a Gaussian line profile (usually a good approximation for the H$\beta$ lines) and use the FWHM to estimate the velocity dispersion.
They use the model-independent method of Peterson, et al. (1998) to estimate the uncertainty in $\tau_{cent}$, so their reported mass errors may be underestimated (as discussed above).

\section{ANALYSIS \& CONCLUSIONS}

\begin{figure}
\figurenum{1}
\plotone{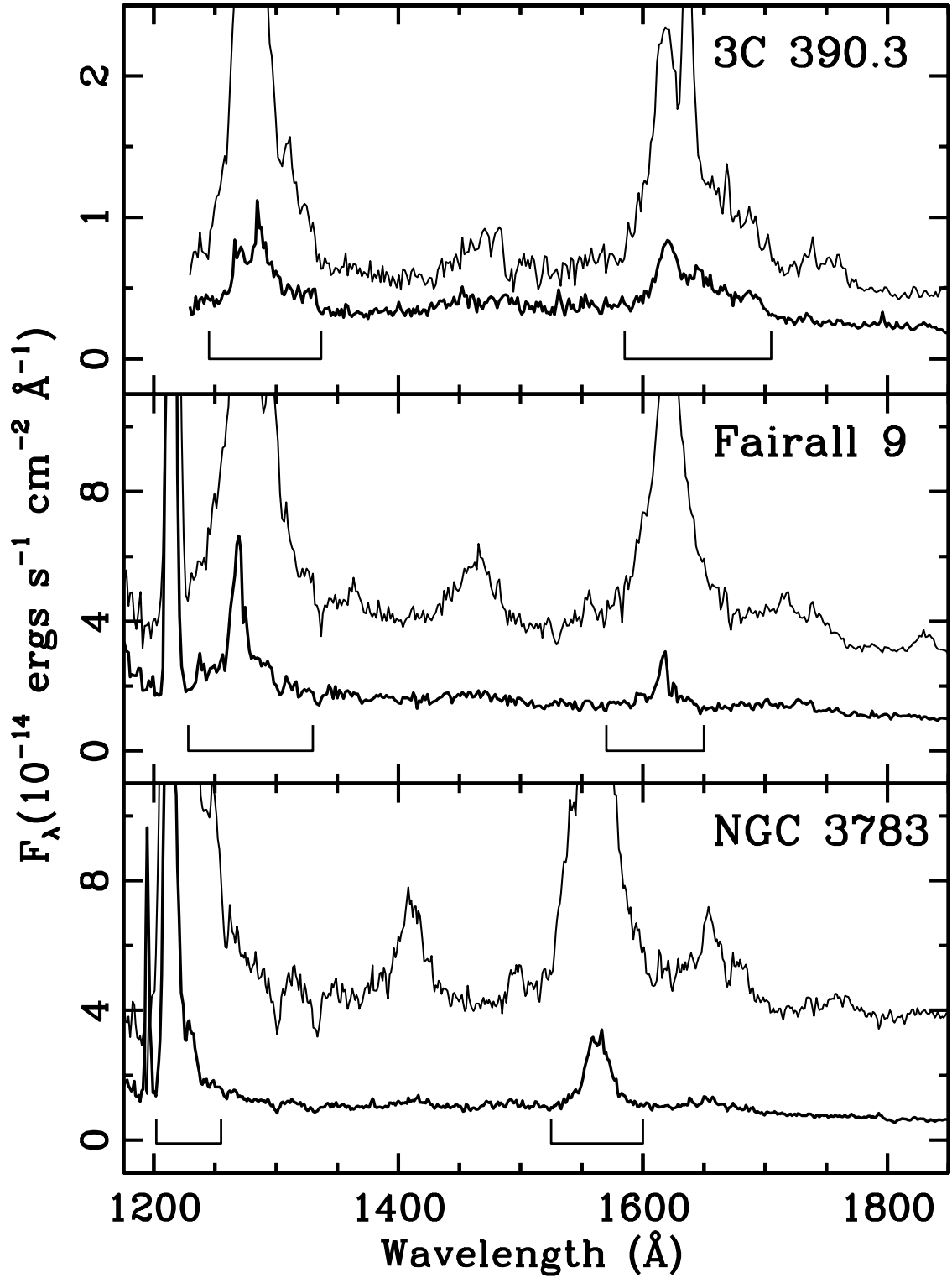}
\caption{Mean (light plot) and rms spectra (dark plot) of 3C 390.3, Fairall 9, and NGC 3783.  The line flux wavelength ranges for  Ly$\alpha$ $+$ N$_{\rm V}$ and C$_{\rm IV}$ are indicated.  Note that, in general, the lines are not well described by Gaussian. \label{fig:3spectra}}
\end{figure}

\begin{figure}
\figurenum{2}
\plotone{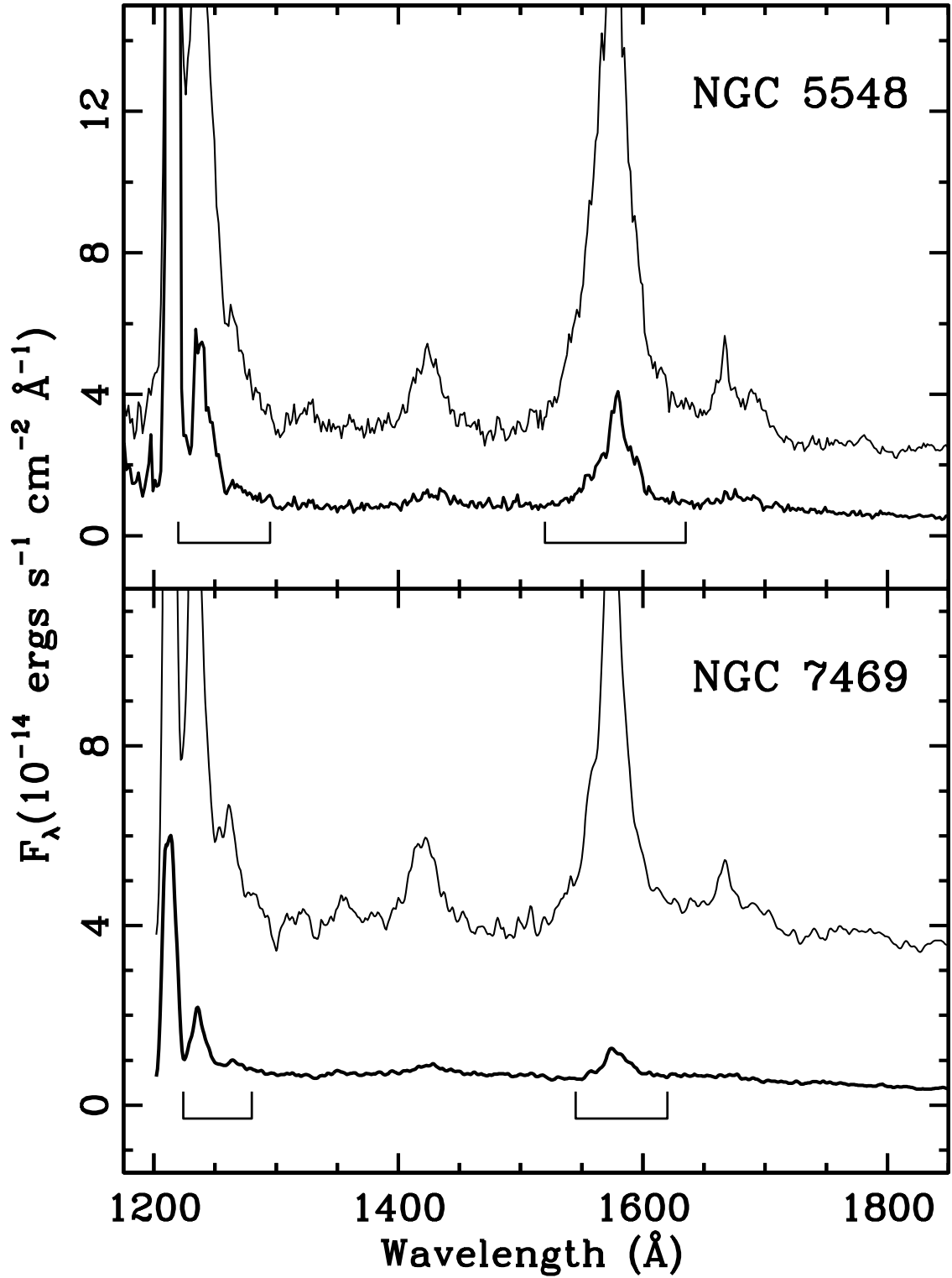}
\caption{Mean (light plot) and rms spectra (dark plot) of NGC 5548 and NGC 7469.  The line flux wavelength ranges for  Ly$\alpha$  $+$ N$_{\rm V}$ and C$_{\rm IV}$ are indicated.  Note that the Ly$\alpha$ line is show significant galactic contamination in both cases. \label{fig:2spectra}}
\end{figure}

Perhaps the most striking inference we can draw from an inspection of Table~\ref{tbl-1}
is that the masses calculated here are considerably lower than those estimated using
older reverberation data.  For example, Koratkar \& Gaskell (1991b) predicted masses of 
$7.3^{+3.5}_{-3.6} \times 10^7 M_{\odot}$, $2.2^{+0.8}_{-0.9} \times 10^8 M_{\odot}$, 
and $4.7^{+1.9}_{-1.9} \times 10^8 M_{\odot}$ for NGC 3783, NGC 5548, and Fairall 9, 
respectively.  The main discrepancy seems to be that the higher (and regularly-spaced) 
sampling rate adopted by the AGN Watch has facilitated the detection of shorter time 
scale variations and cross-correlation lags for these objects.  An additional effect 
stems from our use of the rms spectrum rather than the mean spectrum to estimate 
$V_{\rm FWHM}$ or $\sigma$.  A cursory glance at Figures \ref{fig:3spectra} and 
\ref{fig:2spectra} suggests that the velocity dispersions of the rms spectral lines 
are \emph{smaller} than those of the mean spectra.  The physical interpretation of this effect is not yet clear, though it may be signaling that the emission from the innermost BLR gas does not vary strongly in response to continuum changes.

Our mass estimates agree with the Wandel, et al. (1999) results within error bars in all cases except the Ly${\alpha}$ results of 3C~390.3 and NGC~3783.
The NGC~5548 mass estimate is also consistent with the result obtained by Peterson 
\& Wandel (1999) using the same spectral data, though with the assumption that the
lines are Gaussian.  
Collier et al.~(1998) estimated the mass of NGC~7469 using 
the spectral variability in the optical regime combined with a measurement of $V_{\rm FWHM}$ 
for the  H$\alpha$ and H$\beta$ lines to obtain a mass for this object of 
$8.3 \pm 2.4 \times 10^6 M_{\odot}$, which is again consistent with our result.
These cases provide some support for the validity of our method, since they show
a consistent mass determination for sources that have ``normal'' (i.e., Gaussian) line profiles.
Our method is clearly more general, since it appears to be applicable also to sources
in which the velocity distribution of the BLR does not naturally produce a Gaussian line shape.

A caveat one must consider with this procedure is that the error associated with the Ly${\alpha}$ fit
can be significant.  The mass estimates made using the Ly${\alpha}$ data are higher than 
those based on the \ion{C}{4} line in all cases except Fairall 9. As we discussed above,
this is probably due to a contamination of the line, which suggests that the \ion{C}{4} 
results may be more reliable.  It is possible that this problem may be alleviated by 
line deblending techniques, which we have not incorporated into our analysis here.

Our results support the argument made by Peterson \& Wandel (1999) that a very large
mass must be present within a distance of a few light-days from the centers of many
AGNs.  For the class of $5$ objects in our sample, this mass is at least
$8 \times 10^6\; M_{\odot}$, which is rather compelling evidence for the presence of
a point-like object.  This mass density is even higher than that in the Galactic
Center, where we are reasonably sure that the enclosed mass within $0.015$ pc of the 
nucleus is close to $2.6\times 10^6\;M_\odot$. Alternative models for the latter
that invoke a distribution of dark objects, such as neutron stars, white dwarfs, or
$\sim 10\;M_\odot$ black holes, can now be ruled out on the basis that in equilibrium,
such a distribution would have a core radius significantly larger than the observed
limiting size of $0.015$ pc.  In the sample of AGNs considered here, this exclusion
of a distributed dark matter contribution to the enclosed mass appears to be even
stronger (cf. Maoz 1998).

One of the most important results presented here is the relatively accurate determination 
of the central point mass in these objects.  With the exclusion of 3C 390.3, which appears 
to belong to a different mass class, the masses of the other four AGNs (all Seyfert I's)
are very close to each other and are near the low end of the mass scale for AGNs.
This characteristic, and the fact that the SMBH at the Galactic Center has a mass
within a factor of $4$ or so of this range, hints at a possibly similar formation and
evolutionary history for these nuclei.  In the Galactic Center, the power generated by
the supermassive black hole seems to be the quasi-steady accretion of plasma
ejected by wind-producing stars in the circumnuclear environment, with a rate $\dot M_{gc} 
\approx 2\times 10^{-4}\;M_\odot$ yr$^{-1}$ (Coker \& Melia 1997).  Thus, given a central 
mass $M_{gc}\approx 2.6\times 10^6\;M_\odot$, the characteristic time $M_{gc}/\dot M_{gc}\approx 
13$ billion years suggests that the black hole Sgr A* may have grown to its current
size gradually over a Hubble time via the steady accretion of gas from its environment.
We speculate that the nuclei in the 4 Seyfert I's we have considered here may themselves
have formed and evolved in a similar fashion, and that the range in masses may therefore
be due principally to a variation of the nuclear gas density from which the black holes
accrete.  

\section{ACKNOWLEDGMENTS}
This work was supported by an NSF Graduate Fellowship at the University of Arizona, 
by a Sir Thomas Lyle Fellowship for distinguished overseas visitors at the University 
of Melbourne, and by NASA grant NAG58239.  We thank Bradley Peterson for helpful discussions.

\clearpage
 
\begin{deluxetable}{lllllll}
\footnotesize
\tablecaption{Velocity Dispersions and Mass Estimates of AGNs. \label{tbl-1}}
\tablewidth{0pt}
\tablehead{
\colhead{} & \colhead{}   & \colhead{$\tau_{cent}$}   & \colhead{$\sigma$\tablenotemark{a}} & \colhead{Mass\tablenotemark{b}}  & \colhead{} &\colhead{$M_{W}$\tablenotemark{c}} \nl
\colhead{AGN} & \colhead{Emission Line}   & \colhead{(days)}   & \colhead{(km s$^{-1}$)} & \colhead{($10^7 M_{\odot}$)}  & \colhead{Reference} & \colhead{($10^7 M_{\odot}$)}
} 
\startdata
3C 390.3 ..... & Ly$\alpha$ $\lambda 1240\ + $ N$_{\rm V}$ $\lambda 1240$ & $67^{+19}_{-14}$ & $7290^{+430}_{-490}$ & $70^{+28}_{-24}$ & O'Brien et al. 1998 & $39^{+12}_{-15}$ \nl
 & C$_{\rm IV}$ $\lambda 1549$ & $41^{+14}_{-7}$ & $8110^{+370}_{-430}$ & $53^{+23}_{-15}$ \nl

Fairall 9 ..... & Ly$\alpha$ $\lambda 1240\ + $ N$_{\rm V}$ $\lambda 1240$ & $15^{+7}_{-5}$ & $6180^{+810}_{-530}$ & $11^{+8}_{-6}$ & Rodr\'{i}guez-Pascual et al. 1997 & $8.7^{+2.6}_{-4.5}$ \nl
 & C$_{\rm IV}$ $\lambda 1549$ & $32^{+15}_{-12}\tablenotemark{\dagger}$ & $4380^{+770}_{-1070}$ & $12^{+10}_{-10}$ \nl

NGC 3783 ..... & Ly$\alpha$ $\lambda 1240\ + $ N$_{\rm V}$ $\lambda 1240$ & $4.7^{+2}_{-1}$ & $4140^{+630}_{-930}$ & $1.6^{+0.8}_{-0.4}$ & Reichert et al. 1994 & $1.1^{+1.1}_{-1.0}$ \nl
 & C$_{\rm IV}$ $\lambda 1549$ & $3.9^{+2}_{-1}$ & $4060^{+240}_{-250}$ & $1.3^{+0.8}_{-0.5}$ \nl

NGC 5548 ..... & Ly$\alpha$ $\lambda 1240\ + $ N$_{\rm V}$ $\lambda 1240$ & $6.9^{+3}_{-2}$ & $6520^{+370}_{-430}\tablenotemark{\ddagger}$ & $5.7^{+3.1}_{-2.4}$ & Korista et al. 1995 & $6.8^{+1.5}_{-1.0}$ \nl
 & C$_{\rm IV}$ $\lambda 1549$ & $7.0^{+3}_{-2}$ & $6320^{+570}_{-750}$ & $5.5^{+3.3}_{-2.9}$ \nl

NGC 7469 ..... & Ly$\alpha$ $\lambda 1240\ + $ N$_{\rm V}$ $\lambda 1240$ & $2.3^{+1}_{-1}$ & $5460^{+130}_{-130}\tablenotemark{\ddagger}$ & $1.3^{+0.6}_{-0.6}$ & Wanders et al. 1997 & $0.76^{+0.75}_{-0.76}$ \nl
 & C$_{\rm IV}$ $\lambda 1549$ & $2.7^{+1}_{-1}$ & $3970^{+430}_{-340}$ & $0.83^{+0.49}_{-0.45}$ \nl

\enddata 

\tablenotetext{a}{The errors in $\sigma$ are probably underestimated due to wavelength bin assignment and line contamination.}
\tablenotetext{b}{Mass estimates are subject to systematic error due to our assumption of isotropy.}
\tablenotetext{c}{The mass estimates reported in \cite{AW99} based on H$\beta$ reverberation data.}
\tablenotetext{\dagger}{\cite{PMRP97} attach no significance to this value.}
\tablenotetext{\ddagger}{The Ly$\alpha$ profile in this source is badly contaminated by geocoronal emission.}

\end{deluxetable}

\clearpage

\end{document}